\documentclass[twocolumn,aps,showpacs,floatfix]{revtex4}
\usepackage{graphicx}
\usepackage{dcolumn}
\usepackage{amsmath}
\usepackage{natbib}

\newcommand{\bra}[1]{\langle #1|}
\newcommand{\ket}[1]{|#1\rangle}

\begin{document}
\title{Intermolecular Adhesion in Conjugated Polymers}
\author{Jeremy D. Schmit}
\affiliation{Department of Physics, Brandeis University Waltham, MA 02454}
\author{Alex J. Levine}
\affiliation{Department of Chemistry \& Biochemistry and The California Nanosystems Institute\\
UCLA, Los Angeles, CA 90095-1596}

\date{\today}

\begin{abstract}
Conjugated polymers are observed to aggregate in solution.
To account for this observation we propose a inter-chain binding
mechanism based on the intermolecular tunneling of the delocalized
$\pi$-electrons occurring at points where the polymers cross.
This tunneling mechanism predicts specific bound structures of chain that
depend on whether they are semiconducting or metallic.
Semiconducting chains should form polyacene-like states exhibiting binding at every other
site, while (doped) metallic chains can bind at each site.
We also show that solitons co-localize with the intermolecular
binding sites thereby strengthening the binding effect and investigate
the conformational statistics of the resulting bimolecular aggregates.
\end{abstract}
\pacs{82.35.Cd,83.80.Rd,61.25.Hg}

\maketitle

Conjugated polymers are of great fundamental interest
in that they are exotic, low-dimensional
(semi-) conductors\cite{Su:79,Takayama:80}.
More recently the use of such polymers has been explored in a variety of
applications  ranging from optical \cite{Schwartz:03,Friend:99}
and electronic\cite{Garnier:1994} devices to biosensors\cite{Chen:99}.
While exploring their use as highly sensitive
biosensors, researchers\cite{Wang:01} discovered their surprising
tendency to aggregate into multi-molecular filaments.  This aggregation is even
more surprising in that it occurs in a system of strongly charged
conjugated polymers dissolved in low ionic strength aqueous solutions.

Motivated by this observation, we investigate more
generally a potential aggregation mechanism that should play
a role in any $\pi$-conjugated system. In this letter we propose
that the appearance of intermolecular electronic states made
possible by $\pi$-electron wavefunction
overlap where two conjugated polymers cross leads to a large
enough reduction of the electronic energy of the two polymer
system to cause intermolecular
binding even in dilute solution. The electronic intermolecular
interaction draws localized states out of the continuum of
extended states on each chain in a manner analogous to the
formation of traditional bonding/anti-bonding
orbitals.  We examine the effect of this bonding mechanism in
both semiconducting (SC) and doped metallic (M) polymers.
In the case of the SC polymers we also
examine the interaction of solitonic
excitations on the chains with the binding sites.
Finally we use the results of these
analyses to predict the aggregated fraction of chains in
solution and the change in the polymer conformational
statistics that may be measured via scattering.

\begin{figure}
\includegraphics[width=6cm]{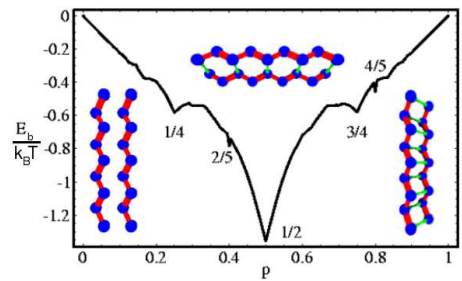}
\caption{(color online) The energy of the
bound undoped chains as a function of the
fraction $p$ of (equally spaced) binding sites for $t'=0.25$eV. The insets show (left to right)
two free chains ($p=0$), and the expected forms of the maximally
bound SC chains at $p=1/2$ and M (doped) chains at $p=1$.
The former structure is similar to polyacene except the binding
results from the hybridization of the $\pi_z$ orbitals.}
\label{binding-pic}
\end{figure}

We find that the inter-chain electronic tunneling leads to a
significant change in the electronic states far from the Fermi level.
In SC polymers, this change makes energetically favorable a polyaclene-like conformation
(see Fig.~\ref{binding-pic}) with binding at every other site.
In addition, there is a strong interaction between these inter-chain binding sites and
solitonic excitations on the chain.
Uncharged solitons are strongly attracted to the interchain binding points and, at
room temperature, co-localize there further strengthening the inter-chain attractive
interaction. Charged solitons, on the other hand, interact weakly with the tunneling
sites at inter-chain crossing points.  In M chains, however, the inter-chain
electronic interaction favors the formation of a parallel polymer configuration where
there is electronic tunneling at every site (see Fig.~\ref{binding-pic}) and
there are no solitonic excitation on the chains.

In either case the presence of
intermolecular bonds changes the statistical properties of the
conformations of the bound pairs primarily by modifying their
effective persistence length. As the binding strength increases, the dominant
deformation mode of the polymers goes from bond rotations in the free chains
to localized bending at defects in the intermolecular bonding pattern for
weakly bound chains, and finally to the elastic bending of stiff polymer bundles.

To consider the most simple description of
conjugated polymers, we study a tight-binding model of noninteracting electrons based
on the SSH Hamiltonian\cite{Su:79}
for each polymer chain. At a specific set of
sites $\alpha$ ({\em i.e.} the crossing points between chains) we allow
inter-chain tunneling with the overlap
integral $t' \simeq 0.2$eV~\cite{Bredas:02}.
Thus, we write for the
single particle Hamiltonian
\begin{eqnarray}
\label{tight-binding}
H&=&\sum_{\ell,n} \left[-t_{\ell,\ell+1}(\ket{\ell,n}
\bra{\ell+1,n}+\ket{\ell+1,n}\bra{\ell,n})\right] \nonumber \\
& -& t'\sum_{\{\alpha\}}\left[\ket{\alpha,1}\bra{\alpha,2}+
\ket{\alpha,2}\bra{\alpha,1}\right].
\end{eqnarray}
Here the $\ket{\ell,n}$ represents an electron on the
$\ell^{\mbox{th}}$ tight-binding site of the $n{^{\mbox{th}}}$ chain
where the sum $\ell = 1,\ldots N$ extends over all sites and $N$ is the
polymerization index and $n=1,2$ labels the two chains. In the case of the
undoped, homogeneously dimerized chain
$t_{\ell,\ell+1}=t_0+(-1)^\ell \frac{\Delta}{2}$. We expect $t
\simeq 2.5$eV and $\Delta \simeq 0.5$eV\cite{Su:79}.
For homogeneous the M chains ({\em i.e.} without the Peierls distortion) $\Delta =0$.

\begin{figure}[htpb]
\includegraphics[width=6.0cm]{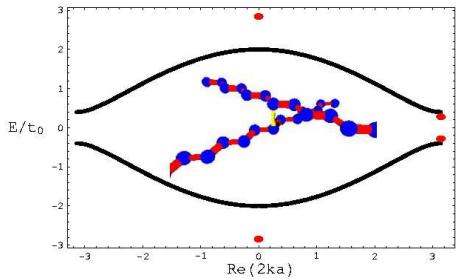}
\caption{(color online) The spectrum of electron states resulting
from electronic intermolecular tunneling at the crossing point. The bound states are
represented by the (red) filled circles at $\mbox{Re}(k)=0,\pi$. The schematic figure
in the center represents such a single tunneling site between two crossing polymers.}
\label{crossmodel}
\end{figure}

We first consider a single crossing point between two homogeneous, infinite,
undoped chains using a transfer matrix
approach\cite{Schmit:06}. The Hamiltonian of the system commutes
with the chain interchange operator so that the energy eigenstates have
definite parity with respect to chain interchange. We find four bound states,
two in each of the symmetric and antisymmetric sectors of the state space, at energies
\begin{equation}
\label{boundstate-energies}
E_b=\pm \sqrt{t_1^2+t_2^2+\frac{t'^2}{2}\pm
\sqrt{4t_1^2 t_2^2+t'^2(t_1^2 +t_2^2)+\frac{t'^4}{4}}},
\end{equation}
where we have defined: $t_{1,2}=t_0 \pm \frac{\Delta}{2}$.
These bound states are represented as red circles on
the band structure diagram shown in Fig.~\ref{crossmodel}. There are two
ultraband~\cite{Phillpot:87} states above
and below the valance and conduction bands. These states are split from the
band edges by  ${\cal O}(t'^2/t_0)$ and their
wavefunction decays over ${\cal O}(t_0/t')$ sites from the tunneling site.
Due to this localization of the bound state wavefunction, two tunneling sites
separated by more than this distance interact weakly. In addition, there are
two states that appear symmetrically
shifted by ${\cal O}(t'^2\Delta/t_0^2)$ into the gap. They have a
real part of their wave number at the edge of the zone and
localization length ${\cal O}( a t_0^2/t' \Delta)$. The effect of the polymer
ends on all of these localized states is insignificant unless the tunneling site
is within a localization distance of the chain ends. For high
molecular weight chains, we may ignore end effects.

The dominant part of the reduction of the system's electronic energy is due
to the occupation of the lower
ultraband state; the contribution of the gap state can be neglected. Similarly,
the shifts in the extended states of the chains are small for high molecular
weight chains since these shifts vanish as $O(N^{-1})$.  Previous calculations
on the M chains~\cite{Schmit:05} demonstrate that the
ultraband state is displaced by $2 t [ \sqrt{1 + t'^2/(4 t^2)} - 1]$ below the
bottom edge of the band, while, once again, the effect of the shift in the
continuum states and the gap state are irrelevant. In the case of both the
M and SC chains, we note that the appearance of a single
localized state centered at the crossing point of two polymers
reduces the electronic energy of the chains by $\sim 100$meV.

We now consider the most energetically favorable configuration
of the two chains in both the
M and SC cases. Here we demonstrate a distinct difference between these two cases.
Previous work on the M polymers has shown that the strongest
interchain binding in the absence of chain electrostatic repulsion
occurs for the parallel chain configuration as shown on the right
side of Fig.~\ref{binding-pic}.  In this case the individual
localized ultraband states below the band edge hybridize into a new
filled band. As this result seems intuitively reasonable, it is
surprising to find that in the case of the SC chains the parallel
configuration of the two chains results in {\em no binding energy} at all.

One can understand this absence of intermolecular binding in the
parallel configuration as follows. In this
configuration the inter-chain interaction symmetrically splits the
filled valance band of the SC polymer. Since the band remains
well-separated from the similarly split conduction band there is no
net energy reduction upon splitting.  For the case of strong enough
inter-chain tunneling ($t' \simeq \Delta$) so that the valance and
conduction bands do not remain well separated, the Peierls distortion
of the chains breaks down and returns the system to the case of
M polymers\cite{Bredas:02}. Instead we find that the energy
minima for SC chains is the polyacene-like structure shown in the
middle cartoon in Fig.~\ref{binding-pic}.

To explore this further we compute the electronic contribution to
the binding energy by direct diagonalization of the two-chain
Hamiltonian with $200$ tight binding sites on each; the results are
shown in Fig.~\ref{binding-pic}. This figure is symmetric about the
$p=1/2$ tunneling site density that is the minimum energy configuration.
The symmetry between
binding site densities $p$ and $1-p$  can be understood as follows.
Removing a single binding site from the fully linked chains can be
represented by adding an inter-chain tunneling term to
Eq.~\ref{tight-binding} with $t' \rightarrow - t'$. Thus a single
non-tunneling defect in the fully linked chains generates the same
set of localized states as does an isolated tunneling site on the
unlinked chain. From Eq.~\ref{boundstate-energies} both localized
states lower the electronic free energy of the system by the same amount.
We extend this argument to multiple non-tunneling defects in the otherwise
fully linked chains by noting that the
associated localized states simply hybridize to a degree that
depends on their spatial separation. At any finite density of
regularly spaced tunneling (non-tunneling) defects the localized
states form a narrow ``impurity'' band, but the symmetry between the
two cases is unchanged. The effect of quenched spatially randomness
in the distribution of the tunneling (non-tunneling) sites on the
energy of the system is minimal~\cite{Schmit:05}.

\begin{figure}
\includegraphics[width=6.0cm]{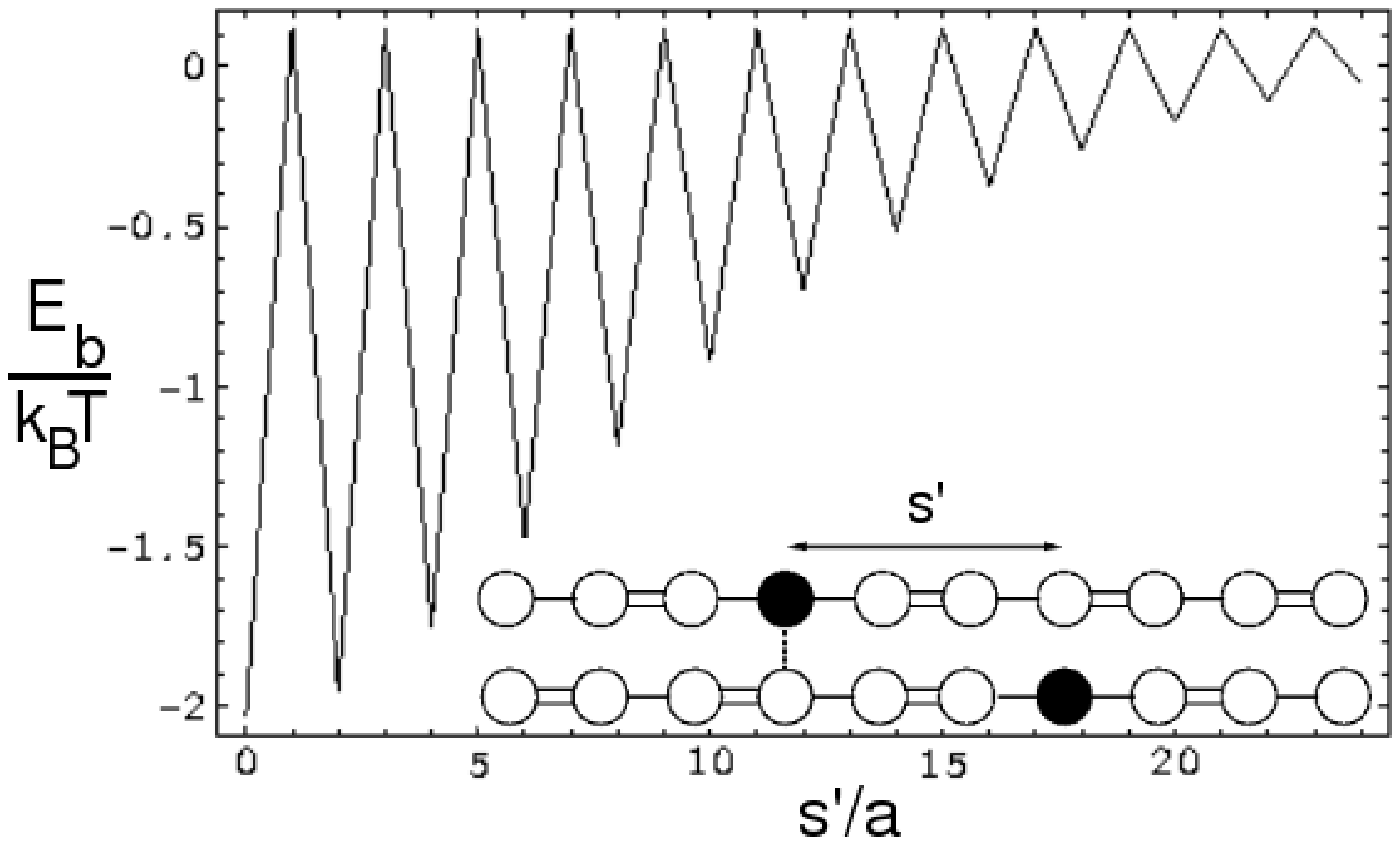}
\includegraphics[width=6.0cm]{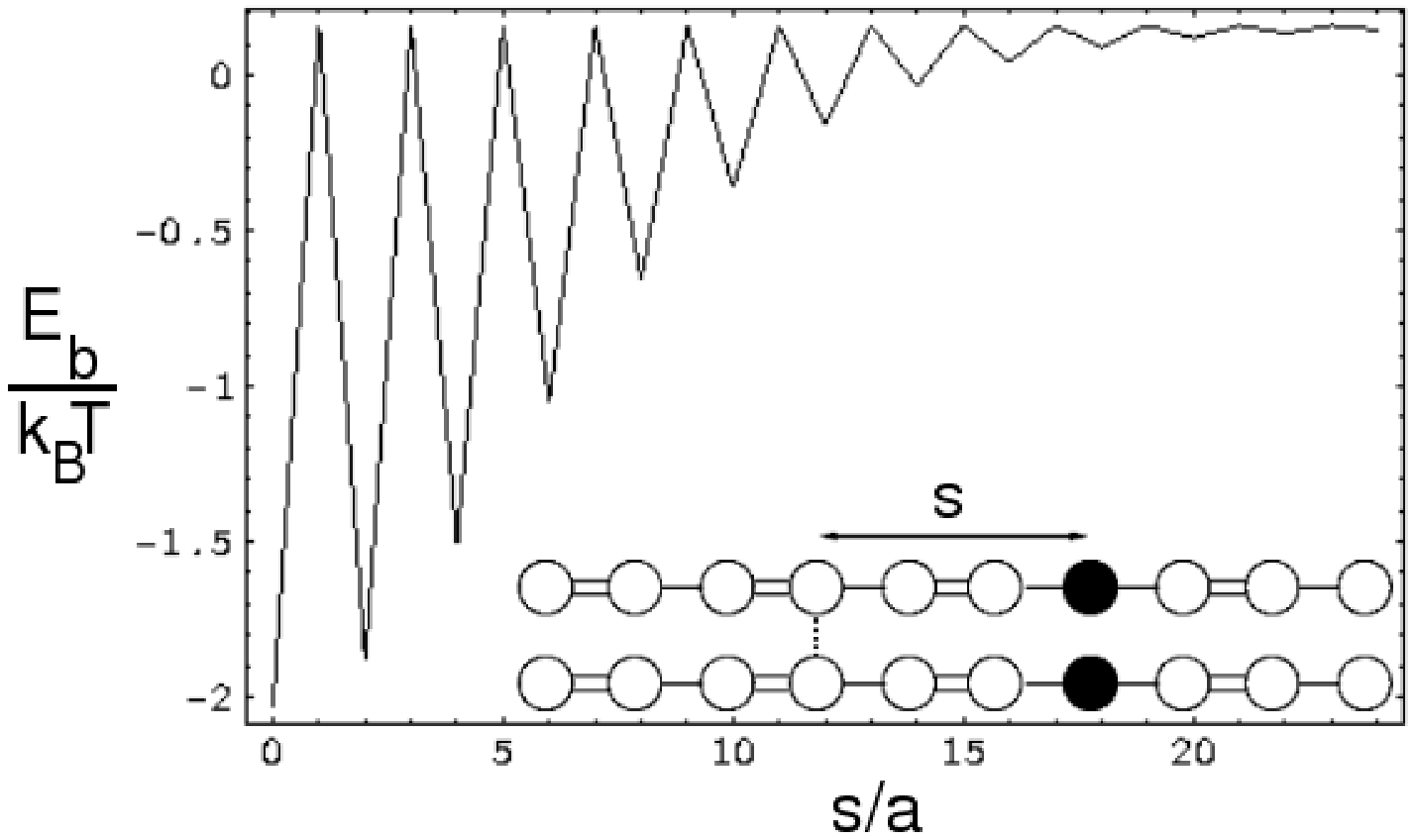}
\caption{The upper figure shows the interaction energy of
two solitons one on each chain separated by $s'$ lattice sites. One soliton
is centered at the binding site. The lower figure shows the interaction energy
of a pair of symmetrically placed solitons displaced from the binding site by $s$.}
\label{soliton-binding}
\end{figure}

In SC chains the inter-chain binding interaction is complicated by
disorder in the dimerization pattern that takes the form of  solitons.
These defects may arise due to either end effects, {\em cis-} insertions
in the {\em trans-} chain, or
light electron/hole doping\cite{Su:79,Roth:87}. We now explore the interaction
between these solitons and the binding sites on the two chains by
considering two chains linked at one tunneling site and examining
the energy spectrum  of the system as function of the relative
positions of the solitons from the binding site.

First, we posit that each chain contains a single
soliton and vary the position of a
the inter-chain tunneling site with respect to the positions of
the soliton. See the inset of Fig.~\ref{soliton-binding}b for
schematic diagram. In moving the tunneling site, we
require the tunneling site to remain on equivalent sites
on the two chains to preserve the simultaneous diagonalization of the
Hamiltonian and the chain interconversion operator. Initially,
we approximate the soliton as a single-site shift in the dimerization.
We consider a more physical model of a
spatially extended soliton below, but this simplified model
captures the basic effect of the solitons in an analytically tractable manner.

\begin{figure}
\includegraphics[width=6cm]{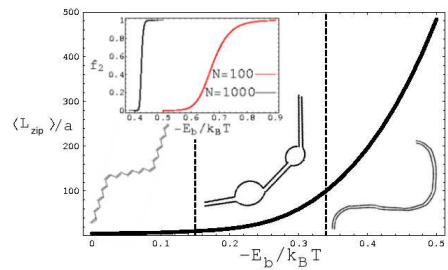}
\caption{(color online) The mean length $\langle L_{\rm zip}
\rangle$ of the bound regions versus the binding $E_b$ energy between
tunneling sites on the two chains. The inset shows the fraction
$f_2$ of bound chains as a function of $E_b$ in dilute solution.}
\label{Lzipfig}
\end{figure}

The single site soliton introduces new localized states at
energies $E$ given by the roots of the polynomial,
\begin{equation}
\label{polynomial-soliton}
0=E^3-E(4[t_0^2+\Delta^2]+t'^2)\pm 8 t' \Delta,
\end{equation}
where the upper (lower) sign applies to the chain
anti-symmetrized (symmetrized) states.  For large $E$ we find
bound states at $ E \simeq (4[t_0^2+\Delta^2]+t'^2)^{1/2}$
similar to the previously discussed ultraband states.   For small
$E$, however, we find a pair of gap states with energy $E \simeq \pm t' \Delta/t_0$
that are symmetrically positioned about
the mid-gap of the SC chain. These are remnants of the usual
soliton associated mid-gap states of the two chains split by the
tunneling matrix element $t'$. The mid-gap state associated with
an uncharged soliton on the chain is singly occupied so that the
splitting of these states allows for the double occupancy of the
lower state resulting in a reduction of  the system's energy upon
the creation of the inter-chain tunneling site. The localization
of an uncharged soliton on each chain at their crossing point further
lowers their electronic energy by $\Delta/t' \simeq 2.5$ enhancing
the binding energy at that site. In contrast, for the
case of charged solitons (as would occur on lightly doped,
SC polymers) both mid-gap states on each chain are filled
there is no further energy reduction due to the
co-localization of the charged solitons with the binding site.

To extend this result to more physical solitons we put
one SSH soliton of width $7a$ at the center of each chain
and numerically calculate the binding energy for the $N=199$ chains.
The interaction energy between the SSH solitons and the tunneling
site is shown in Fig. \ref{soliton-binding}.  We find that
the solitons are attracted to the inter-chain tunneling site.
The width of this potential well for solitons is broadened
once one soliton reaches the binding site. The interaction
potential between the solitons and the binding site is oscillatory
because the soliton associated mid-gap states have nodes on alternating sites.

The formation of these bimolecular assemblies will effect the
photophysics of the system\cite{Daniel:05} as well as the
conformational statistics of the polymers in solution.  As the site
binding energy $E_b$  is increased there is an abrupt condensation
of the polymers in dilute solution
into bimolecular filaments. See the inset of Fig.~\ref{Lzipfig}.
We do not consider aggregation into larger supramolecular assemblies.
The persistence length of the bimolecular filaments has three distinct
regimes as a function of $E_b$. In the limit of a vanishing $E_b$,
conformations of the free chains may be approximated as a freely jointed
chain with a Kuhn length of $l_{\rm min} \sim 10a$\cite{Gettinger:94}
where $a$ is the length of a unit cell. For $E_b \gg k_{\rm B} T$ the
bimolecular filaments saturate their available binding sites,
and the intermolecular bonds hinder bond rotations so that the softest
deformation mode involves bending in the conjugation plane. Using
known bond bending and stretching
constants\cite{Su:79,Canales:03}, we estimate that the persistence
length is $\ell_{\rm max} \sim 100a$. Between these extremes, where
defects in the intermolecular binding are typically separated by less
than  $\ell_{\rm max}$, the chain tangents decorrelate primarily due
to the essentially free rotations at those defects. Thus the bimolecular
filaments will have a continuously varying persistence length $\ell_{\rm P}$,
$\ell_{\rm min} < \ell_{\rm P} <
\ell_{\rm max}$ and their effective persistence length is determined by the
mean length of the fully bound chain segments
$\langle L_{\rm zip}\rangle$.

We solve for this mean length $\langle L_{\rm zip}\rangle$ using a
modified Poland-Scheraga (PS) model\cite{Poland:70}.  The partition
function for the partially bound pair of chains each containing $N$
sites may be written as
\begin{equation}
\label{parition}
Z(N)=\sum_{p}\sum_{\{i_\pi^{(1)},i_\pi^{(2)},j_\pi\}}\prod_{\pi=0}^p
u(i_\pi^{(1)},i_\pi^{(2)})v(j_\pi),
\end{equation}
where $u(i_\pi^{(1)},i_\pi^{(2)})$ is the Boltzmann weight of the $\pi^{\rm th}$
unbound region consisting of
$i_\pi^{(1)}+i_\pi^{(2)}$ tight binding sites where the superscript
indexes the two chains. $v({j_\pi})$ is the corresponding Boltzmann
weight associated with a bound region of
$j_\pi$ consecutive sites. The sum in Eq.~\ref{parition} is
constrained by the fixed total number of tight binding sites.
We relax this constraint using the Grand Canonical ensemble
and, from the PS technique we compute the free energy
$F = -k_BT N \ln(x_1)$ where $x_1$ is the root of
\begin{equation}
\frac{x(x-v)}{xv-v^2(1-\sigma)}=\sum_{n=1}^\infty e^{- \pi \ell_{min}/n}
\left(\frac{u}{x}\right)^n \left(\frac{n}{2}\right)^{-c}.
\end{equation}
Here $v = \exp(-E_b/k_{\rm B} T)$ is the Boltzmann
weight of a single binding site, while $u \simeq \exp(a/\ell_p) \approx 1$
is the exponentiated entropy of a segment of unbound chain.  The exponent
$c\simeq 2.1$ reflects the loss of entropy in these
unbound regions due to self avoidance and the loop closure
condition~\cite{Kafri:00}. Finally, $-k_BT\ln\sigma$ is the domain
wall energy between bound and unbound regions accounting for local
chain bending energy. The fraction of binding sites is given by
$\theta=\partial \ln x_1/\partial \ln v$ and the mean length of
bound regions $\langle L_{\rm zip}\rangle$ may be computed from
the ratio of $\theta$,  to the fraction of sites at the  boundary
of bound and unbound regions, $\theta_b$. The results are shown
in Fig.~\ref{Lzipfig}.  For $E_b <  0.4 k_{\rm B} T$,
$\langle L_{\rm zip}\rangle <  \ell_{\rm max}$. Thus,
given the best estimate of $t'$ and for $N> 10^3$, chains
will aggregate in dilute solution ($10^{-5}$M)  into
this intermediate state where the persistence length is
continuously controlled by $E_b$. Measurements of the
persistence length may test the estimates of $t'$ arising from quantum
chemical calculations, however, there are other contributions
to $E_b$ including the electrostatic repulsion
between {\em e.g.} the highly charged, water soluble polymers.
More direct tests of this theory necessitate measurements on more chemically simple
$\pi$-conjugated systems such as polyacetylene.

Understanding the persistence length of bimolecular
filaments may play a role in controlling the
penetration of such polymers into microporous or zeolitic
structures. Future work will explore conjugated gels
formed at higher polymer concentrations.

JS was supported by the MRSEC Program of the National Science
Foundation under Award No DMR00-80034.  The authors would like to
thank Fyl Pincus for valuable discussions.

\end{document}